\documentclass[vecphys]{svmult}

\usepackage{makeidx}         % allows index generation
\usepackage{graphicx}        % standard LaTeX graphics tool
\usepackage{multicol}        % used for the two-column index
\usepackage[bottom]{footmisc}% places footnotes at page bottom
\makeindex             % used for the subject index
\begin{document}

\title*{Science with a 16m VLT: the case for  variability of fundamental constants}
% Use \titlerunning{Short Title} for an abbreviated version of
% your contribution title if the original one is too long
\author{Paolo Molaro\inst{1} }
% Use \authorrunning{Short Title} for an abbreviated version of
% your contribution title if the original one is too long
\institute{INAF-OAT, Trieste Via G.B. Tiepolo 11, I 34143 Italy
\texttt{molaro@oats.inaf.it}}
%\and Name and Address of your Institute \texttt{name@email.address}}
%
% Use the package "url.sty" to avoid
% problems with special characters
% used in your e-mail or web address
%
\maketitle

\section{Abstract}
Only astronomical observations  can effectively probe in space-time the variability of the physical dimensionless constants such as the fine structure constant  $\alpha$ and proton-to-electron mass ratio, $\mu$,   which are related to  fundamental forces of nature. Several theories beyond the Standard Model (SM) allow fundamental constants to vary, but they cannot  make quantitative predictions so that only laboratory experiments  and astronomical observations  can show if this is the case or set the  allowed bounds.  At the moment of writing there are claims for a variability of both $\alpha$ and $\mu$   at  5 and 4 $\sigma$ of C.L., respectively,  although for $\alpha$ they are  contrasted by null results. The observations are challenging and a  new spectrograph such as ESPRESSO at the combined  incoherent  focus of 4 VLT units (a potential 16 m equivalent telescope) will allow for a significant  improvement in the precision measurement   clearing up the  controversy.  If the variations will be confirmed,   the implications are  far reaching,  revealing  new physics beyond the SM  and pointing a direction for  GUTs theories.  A most exciting possibility is that a variation of $\alpha$  is  induced by quintessence through its coupling with the  electromagnetic field.  If this is the case  an accurate  measurement of the variability    could provide a way  for   reconstructing the  equation of state of Dark Energy \cite{a06}.
%\label{sec:1}
% Always give a unique label
% and use \ref{<label>} for cross-references
% and \cite{<label>} for bibliographic references
% use \sectionmark{}
% to alter or adjust the section heading in the running head
%\cite{monograph}.

\section{Introduction}

The Standard Model (SM) of particle physics needs  26 dimensionless physical constants for the description of the natural world (\cite{t06}), of these few  are  directly related to the strength of fundamental forces. Among them  the fine structure  constant ($\alpha = {\rm e}^2/(\hbar c) $) and the proton-to-electron  mass ratio, ($\mu  = m_p/m_e$)   are  of particular interest for us since they  can be measured accurately by astronomical observations  of  intervening absorption systems towards distant QSOs. The fine structure constant  $\alpha$ is related to the strength of the electromagnetic force;   $m_e$ is related to the vacuum expectation value of the Higgs
field, namely the scale of the weak nuclear force, and  m$_p$ is related to the $\Lambda_{QCD}$  or the strong nuclear force, therefore  $\mu$ is related to the ratio between the strong and weak nuclear forces.

A whatever small variability  of these constants  will produce a violation of the   Weak Equivalence Principle (WEP) and would have far reaching implications revealing new   physics beyond the Standard Model.  In the Gev  energy regime $\alpha$ has already been shown to vary, but at low energies  laboratory measurements with cooled atomic clocks failed to detect variations at the fifteen decimal place. The  most stringent laboratory value is $\dot{\alpha}/\alpha$  = (-2.7$\pm$3.9)$\cdot 10^{-16}$ yr$^{-1}$  \cite{p06}. Astronomy is providing some evidence for  both $\alpha$ and $\mu$ variations, although the evidence  for $\alpha$ has   been contrasted by  other groups.    
The astronomical  claims  for a variability  are at the level of  6 ppm,  part-per-million, and are measured up to redshift 4, or 12 Gyr  lookback time. 
%Assuming a linear  variability, which might not be the case,  then the  laboratory  bounds  correspond to a bound of 0.5 ppm which starts to be in 
%contradiction with the astronomical evidences.
Several space-based missions as ACES, $\mu$SCOPE, STEP will soon improve existing laboratory bounds for WEP  up to 6 orders of magnitude, and   they should  find violations if present claims of variability are correct under simple linear extrapolation. It is thus desirable that the astronomical community will be able to clear up the case before these accurate  experiments  will fly, but only astronomical observations can  probe   WEP non locally.

\subsection{Why constants should vary?}

Strings and multidimensional theories   predict variable constants since the constants are defined in the whole multidimensional space and vary as  extra dimensions are  varying.  The coupling between  a scalar field 
with the electromagnetic field  gives also varying constants. The 
required cosmological constant value is so small
that a quintessence  is a likely candidate for Dark Energy. Thus varying constants could  provide insights into  the nature of dark energy and provide evidence for scalar fields  \cite{ma07,f07}.
%These are often invoked in the SM as the Higgs field or Inflation theory but in fact we do not know if they exist.
Avelino et al. (2006) have shown how  a  precise detection of the variability of a constant could be used for  the reconstruction of the quintessence potential and of the equation of state of Dark Energy \cite{a06}. 

If one constant is varying, then all the gauge and Yukawa couplings are also expected to vary. There is precise  relation between the variation of $\alpha$ and $\mu$,  but it  depends on  the  context the unification is realized in. Thus, simultaneous measurements of the variability of $\alpha$ and $\mu$ at similar redshift  will be a  key discriminant of the several GUTs models  \cite{d07}. Theoretical preferences are for  a relative change between the $  \mu$ and $\alpha $  variations of $\le$ 50, but larger values are also possible, implying that  the strong-coupling constant is running faster than $\alpha$ and therefore $\delta \mu$  should
be found to be larger then $\delta \alpha$. 

\section{The observations}

Observations of  the Werner and Lyman series of the molecular hydrogen in Damped Ly$\alpha$ galaxies (DLA) can be used to bound $\mu$ variations. The electron-vibro-rotational transitions have different dependence from the reduced mass and can be used to constrain a variability of $\mu$.
UVES  observations of the DLA   at z$_{abs}$ =3.0 towards QSO 0347-383  \cite{i05}, but see  \cite{l02},  and of the DLA  towards QSO 0405-443   have provided $\delta \mu
= (24 \pm 6)$ ppm, when the two systems are combined  together \cite{r06}. The handful of systems investigated for this purpose reflects the difficulties of the measurement. There are few DLA showing H$_2$  and  the restframe H$_2$  lines are at $\approx$ 1000 \AA\,,  falling in the Lyman forest and   requiring a z$_{abs} \ge$ 2  to be redshifted  into the optical window.  H$_{2}$ systems are  extensively searched at the moment so  that probably new observations will be available in the near future  to verify  these first findings.

Fine structure  variability can be probed in the early universe through the primordial nucleosynthesis or through the CMB power spectrum but at the level of a few percent. The most effective way  has been achieved through the analysis of metal lines of intervening absorption systems observed in the spectra of distant  QSOs.  
The energy  levels of high mass nucleus are subject to  relativistic corrections which are sensitive to the  mass number.
These have  been calculated for the most frequently observed resonance lines  and constitute the   popular      Many-Multiplet method. 
Murphy and collaborators  \cite{mu04}
 by comparing the redshift of several lines in  a sample of 143 systems in the redshift interval 0.2$<z_{abs}<$4.2    found evidence for
$\Delta \alpha / \alpha$ = $(-5.7\pm 1.1)$  ppm.  However, this evidence has been contrasted by two other groups which did not find
evidence for variability at the level claimed. Chand et al.  found an average value of  $(-0.6 \pm 0.6)$ ppm in a sample of 23 systems, while  Levshakov and collaborators  found $(-0.12 \pm 1.79)$  ppm and $(5.66 \pm 2.67)$ ppm in two systems at  $z = 1.15$ and 1.84, respectively, and  by using lines of Fe\,{\sc ii}  only \cite{q04,l07}. What is the best  methodology  is currently under   debate \cite{mu07a,mu07,s07,m07}.
\begin{figure}
\centering
% Use the relevant command for your figure-insertion program
% to insert the figure file.
% For example, with the option graphics use
\includegraphics[height=5cm]{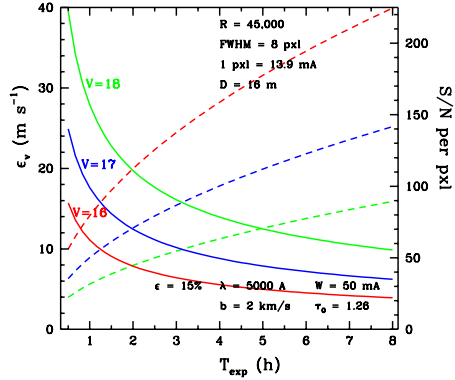}
%
% If not, use
%\picplace{5cm}{2cm} % Give the correct figure height and width in cm
%
\caption{Estimated accuracy in the position of an absorption line with EW =0.050 \AA\ and $b$= 2 km s$^{-1}$ as a function of the exposure time for ESPRESSO@4UT. On the figure legend   other instrumental and observational parameters are given. }
\label{fig:1}       % Give a unique label
\end{figure}

\subsection{Would you like an ESPRESSO?}

These observations are  challenging the instrumental performances of UVES-VLT or HIRES-Keck telescopes.
Measuring the variability of  $\mu$ or $\alpha$ implies the  measurement  of a tiny variation of the position of one or few lines with respect to  other reference lines.
It is not much different than revealing exoplanets, but with the limitations  that only few lines can be used and QSO are much fainter than stellar sources.
The precision in the  measure of a line position increases  with  the spectrograph resolving power till the intrinsic broadening of the  metal lines is resolved, the signal-to-noise and with the decreasing of the pixel size
($\Delta \lambda$ $^{3/2}$, see  \cite{b83} for a precise relation). 
The ESPRESSO spectrograph described by L. Pasquini at this conference, both in  the {\it Super-HARPS} or  {\it Super-UVES}  modes, holds the promise for   one order of magnitude  improvement   compared to  what presently achieved.  Fig. 1  shows the accuracy which can be achieved in the photon limit approximation and accuracies of few 10 ms$^{-1}$ are  reachable for single lines with relatively short exposures even for faint sources. An error of 30 m s$^{-1}$  corresponds  to an error of 1 ppm for $\alpha$; 
such an accuracy will be  enough to resolve the present controversy and establish in a definitive way whether $\alpha $  or $\mu$ are varying  as claimed. However, one important requirement is  the   improvement of the  wavelength calibration, for instance with the LaserComb as  discussed here by A. Manescau.

 \section{Constants and Dark Energy}

 Avelino and collaborators  \cite{a06} have shown that  the measurement  of the behaviour of  variations in $\alpha$  and  $\mu$ with redshift can be used to infer the evolution of the scalar field and of the equation of state  of the Dark Energy,  not very differently from   the reconstruction of the potential from the motion of a particle.  Nelson Nunes kindly  adapted  their detailed analysis to a realistic set of observations which can be performed with ESPRESSO@4VLT.  It is assumed that it has been possible to measure $\alpha$ and $\mu$ for a sample of 200  and  25 systems respectively and with an equal, for simplicity,  accuracy of 1 ppm.
In the example case the scalar potential is taken as V($\phi$) = V$_0$(exp(10k$\phi$ + exp(0.1k$\phi$)), which is one of the simplest possible potential accounting for the accelerated expansion.  Fig. 2 shows the Monte Carlo redshift distribution of the data with this scalar potential assuming that the variation of $\alpha$ {\bf is} -5 ppm at z=3 and  that  the two constants are mutually linked by a fix ratio of -6, as it is suggested by some of the  observations. In Fig. 3  the red dotted line shows the assumed behaviour of the {\it w(z)} while the black continuos line shows its  recovering  through a  fitting of the simulated data points with a polynomial of order m=3 (cfr  \cite{a06} for details). The shaded regions show the 1 and 2 CL  of the reconstruction, when  both  $\alpha$  and $\mu$ measurements have been considered. We emphasize that only few observations would clearly show if  {\it w(z)} is an evolving function of z.

\begin{figure}
\centering
% Use the relevant command for your figure-insertion program
% to insert the figure file.
% For example, with the option graphics use
\includegraphics[height=5cm]{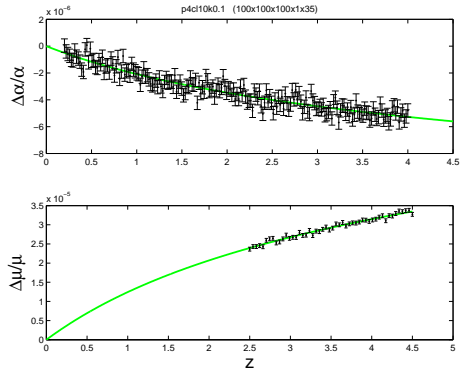}
%
% If not, use
%\picplace{5cm}{2cm} % Give the correct figure height and width in cm
%
\caption{Monte Carlo data set based on redshift
dependence of the scalar potential given in the text producing a $\Delta \alpha / \alpha$ =-5 ppm at z=3. Error bars are of 1 ppm for $\alpha$ 
and $\mu$ as expected with ESPRESSO@4VLT. }
\label{fig:2}       % Give a unique label
\end{figure}

\begin{figure}
\centering
% Use the relevant command for your figure-insertion program
% to insert the figure file.
% For example, with the option graphics use
\includegraphics[height=5cm]{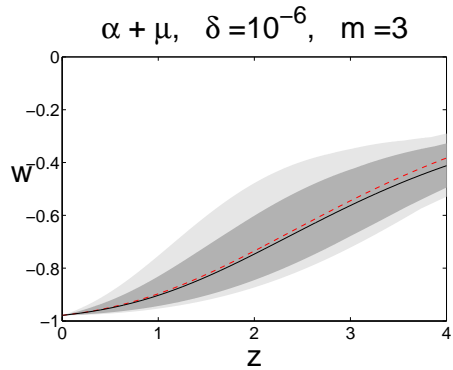}
%
% If not, use
%\picplace{5cm}{2cm} % Give the correct figure height and width in cm
%
\caption{Reconstruction  of the equation of state and its error band. Dashed line represents the assumed dark energy and the solid line the reconstruction's best fit. Shaded regions show the 1 and 2 CL  of the reconstruction}
\label{fig:2}       % Give a unique label
\end{figure}

\section{Conclusions}
Variability of physical constants is an important issue for physics
and only astronomy can probe this possibility  for $\alpha$ and $\mu$ in the full space-time.
 Present observations provide hints of variation for both constants but those for $\alpha$ have been contrasted by other investigations.
The ESPRESSO spectrograph presently conceived for the incoherent combined focus of the 4VLT would  improve present accuracy by a significant factor 
  and therefore clarify the case. A confirmation of the variability would have far reaching  implications revealing   new physics beyond the SM, showing the right path for GUTs and possibly  providing insights into the  nature of Dark Energy.
If no variability is found, then the new more stringent bounds will be usefully combined with local space experiments for WEP violation.
Overall, this seems to be  a great opportunity  for the astronomical community and I hope that ESO will  take advantage of it by considering the construction of the new high precision spectrograph at the incoherent combined focus of the 4 VLT units,  a $\approx$ 16m equivalent telescope.
   % For figures use
%

\section{Acknowledgements}
It is a pleasure to thank N. Nunes for adapting  his simulations for ESPRESSO, all the ESPRESSO collaboration,  in particular S. Levshakov and M. Murphy.
 %
%
% BibTeX users please use
% \bibliographystyle{}
% \bibliography{}
%
% Non-BibTeX users please follow the syntax
% the syntax of "referenc.tex" for your own citations
%\input{referenc}
%%%%%%%%%%%%%%%%%%%%%%%% referenc.tex %%%%%%%%%%%%%%%%%%%%%%%%%%%%%%
% sample references
% "physics"
%
% Use this file as a template for your own input.
%
%%%%%%%%%%%%%%%%%%%%%%%% Springer-Verlag %%%%%%%%%%%%%%%%%%%%%%%%%%

%
% BibTeX users please use
% \bibliographystyle{}
% \bibliography{}
%
% Non-BibTeX users please use

%%%%%%%%%%%%%%%%%%%%%%%%%%%%%%%%%%%%%%%%%%%%%%%%%%%%%%%%%%%%%%%%%%%%%%  }

%%%%%%%%%%%%%%%%%%%%%%%%%%%%%%%%%%%%%%%%%%%%%%%%%%%%%%%%%%%%%%%%%%%%%%

\printindex
\end{document}